\documentclass[twocolumn]{aastex61}

\usepackage{amstext}
\usepackage{amsmath}
\usepackage{amssymb}
\usepackage{graphicx}

\newcommand{\beq}{\begin{equation}}
\newcommand{\eeq}{\end{equation}}

\shorttitle{}

\shortauthors{}%

\begin{document}

\title{Planetary Engulfment in the Hertzsprung--Russell Diagram}

\correspondingauthor{Morgan MacLeod}
\email{morgan.macleod@cfa.harvard.edu}

\author{Morgan MacLeod}
\affiliation{Harvard-Smithsonian Center for Astrophysics, 60 Garden Street, Cambridge, MA 02138, USA}
\altaffiliation{NASA Einstein Fellow}

\author{Matteo Cantiello}
\affiliation{Center for Computational Astrophysics, Flatiron Institute, 162 5th Avenue, New York, NY 10010, USA}
\affiliation{Department of Astrophysical Sciences, Princeton University, Princeton, NJ 08544, USA}

\author{Melinda Soares-Furtado}
\affiliation{Department of Astrophysical Sciences, Princeton University, Princeton, NJ 08544, USA}

\begin{abstract}
Planets  accompany most sun-like stars. The orbits of many are sufficiently close that they will be engulfed when their host stars ascend the giant branch.  
This Letter compares the power generated by orbital decay of an engulfed planet to the intrinsic stellar luminosity.
Orbital decay power is generated by drag on the engulfed companion by the surrounding envelope.  As stars ascend the giant branch their envelope density drops and so does the power injected through orbital decay, scaling approximately as $L_{\rm decay} \propto R_*^{-9/2}$. Their luminosity, however, increases  along the giant branch. These opposed scalings indicate a crossing, where $L_{\rm decay}= L_*$. We consider the engulfment of planets along isochrones in the Hertzsprung--Russell (H--R) diagram. We find that the conditions for such a crossing occur around $L_*\approx 10^2$~$L_\odot$ (or $a\approx 0.1$~au) for Jovian planetary companions.  The consumption of closer-in giant planets, such as hot Jupiters, leads to $L_{\rm decay}\gg L_*$, while more distant planets such as warm Jupiters, $a\approx 0.5$~au, lead to minor perturbations of their host stars with $L_{\rm decay} \ll L_*$.  Our results map out the parameter space along the giant branch in the H--R Diagram where interaction with planetary companions leads to significant energetic disturbance of host stars. 
\end{abstract}

\section{Introduction}\label{sec:introduction}

Planetary companions to extrasolar stars have been uncovered in great abundance by transit and radial velocity searches \citep{2015ARA&A..53..409W}. 
The discovery of numerous giant planets in compact orbits beginning with 51 Peg b  \citep{1995Natur.378..355M} was unexpected given the architecture of our own solar system \citep{2012ApJS..201...15H,2012ApJ...753..160W}. 
About 10\% of approximately solar-mass stars host giant planets with orbital periods of a few years or less \citep{2015ARA&A..53..409W}. Of these, some are in very compact orbits of days (hot Jupiters, $a\lesssim 0.1$~au), while others populate longer orbital periods (warm Jupiters, $0.1$~au$\lesssim a\lesssim1$~au).  More massive substellar companions like brown dwarfs are found to be nearly an order of magnitude less common than giant planets at these orbital separations \citep{2006ApJ...640.1051G}.

As stars evolve, their radii grow significantly. Planetary companions with orbital separations less than au scales can be engulfed within the envelopes of their growing giant-star hosts \citep[e.g.][]{2009ApJ...705L..81V,2013ApJ...772..143S}.  During the ensuing common envelope phase, the orbit of the companion shrinks in response to drag generated by interaction with the surrounding envelope \citep{1976IAUS...73...75P}. 
The deposition of orbital energy is a power source within the stellar envelope. Depending on the relative magnitude of this power source and the nuclear-burning power already transported by radiation and convection through the stellar envelope, this addition may represent a perturbation or a large-scale disturbance. 

A number of authors have considered the impact of digested planets on their host stars. Several  have studied the gas dynamics of planet engulfment \citep{1998ApJ...506L..65S,2016MNRAS.458..832S} or the subsequent stellar evolution \citep[e.g.][]{1984MNRAS.210..189S,1999MNRAS.304..925S,1999MNRAS.308.1133S,2017MNRAS.468.4399M}. 
Others have discussed the role of digested planets in depositing angular momentum into the stellar envelope and enhancing stellar rotation \citep{1998AJ....116.1308S,1999MNRAS.308.1133S,2014ApJ...787..131Z,2016A&A...591A..45P,2016A&A...593A.128P}, enhancing surface magnetic field \citep{2016A&A...593L..15P}, 
and polluting surface abundances with planetary material \citep{1998ApJ...506L..65S,1999MNRAS.308.1133S,2002ApJ...572.1012S,2016ApJ...829..127A,2016ApJ...833L..24A}. 
Finally, some authors have estimated the role of engulfment events in producing transients from stellar ejecta \citep{2006MNRAS.373..733S,2012MNRAS.425.2778M,2017MNRAS.466.1421Y}, or in shaping planetary nebulae from ejecta \citep{2011PASP..123..402D}. 

This work examines one aspect of the broader question of planetary engulfment: the comparison  of orbital decay power to  stellar luminosity. In Section~\ref{sec:orbdecay}, we outline the scalings that determine the rate at which an engulfed object's orbit decays in response to  interaction with a giant star's gaseous envelope. In Section~\ref{sec:hr}, we apply these scalings across stellar isochrones to consider interactions characterized by different star and companion populations in the Hertzsprung--Russell (H--R) diagram. We show that the orbital decay power drops with more extended host stars, while the intrinsic stellar power grows, and highlight the presence of a crossing in the relative importance of these two terms.  In Section~\ref{sec:discussion}, we discuss some implications of our findings in light of previous work, and in Section~\ref{sec:conclusion}, we summarize our findings. 

\section{Orbital Decay Power}\label{sec:orbdecay}

We begin by considering a system composed of a giant star of mass $M_*$, with core mass $M_{\rm c}$, total radius $R_*$, and luminosity $L_*$,  and a lower-mass companion of mass $M_2$ and radius $R_2$ on a circular orbit. Interaction between the star and planet begins when the star fills its Roche lobe and continues while the orbital separation shrinks until the planet is eroded.  During the initial engulfment, the orbital separation is, therefore, similar to the stellar radius. 
When $a=R_*$, the orbital energy is
\beq\label{eorb}
E_{\rm orb} = \frac{G M_* M_2}{2 R_*},
\eeq
the orbital velocity is 
\beq\label{vorb}
v_{\rm orb} =  \left[\frac{G (M_* +M_2)}{R_*} \right]^{1/2},
\eeq
and the associated orbital period is 
\beq\label{porb}
P_{\rm orb} = 2 \pi \left[\frac{R_*^3}{G (M_*+ M_2)}\right]^{1/2}. 
\eeq

The orbit decays in response to a drag force $F_{\rm d}$, directed in opposition to the relative motion between the object and the gas, $v_{\rm rel}$. This force transfers energy from the orbit to the stellar gas at a rate of
\begin{align}
L_{\rm decay} &= F_{\rm d}  v_{\rm rel}, \nonumber \\
&\approx F_{\rm d} v_{\rm orb},
\end{align}
where the second approximation is valid when the surrounding gas is not 
synchronized with the orbital motion ($v_{\rm rel} \approx v_{\rm orb}$, which we adopt below).
The drag force can be expressed in the form  
\beq\label{drag}
F_{\rm d} = C_{\rm d} A \rho v_{\rm rel}^2,
\eeq
where $A$ is a cross-sectional area, $\rho$ is the density of the intervening material, and $C_{\rm d}$ is a dimensionless drag coefficient. 

The drag force experienced by an engulfed object depends on whether the appropriate cross section, $A$, is geometrically or gravitationally determined. The minimum cross section is the geometric, $\pi R_2^2$. This cross section is enhanced by gravitational focusing when the escape velocity of the object, $\sqrt{2GM_2/R_2}$, is larger than the relative velocity.\footnote{In the limit where it dominates, the gravitational focus cross section, $\pi R_{\rm a}^2$, depends on the object's mass (rather than its radius) through the gravitational focusing impact parameter $R_{\rm a}= 2 G M_2 / v_{\rm rel}^2$, \citep{1939PCPS...35..405H}. }
For companions of approximately Jupiter mass and below interacting with stars, the geometric cross section is appropriate and we adopt it here.  

\begin{figure}[tbp]
\begin{center}
\includegraphics[width=0.47\textwidth]{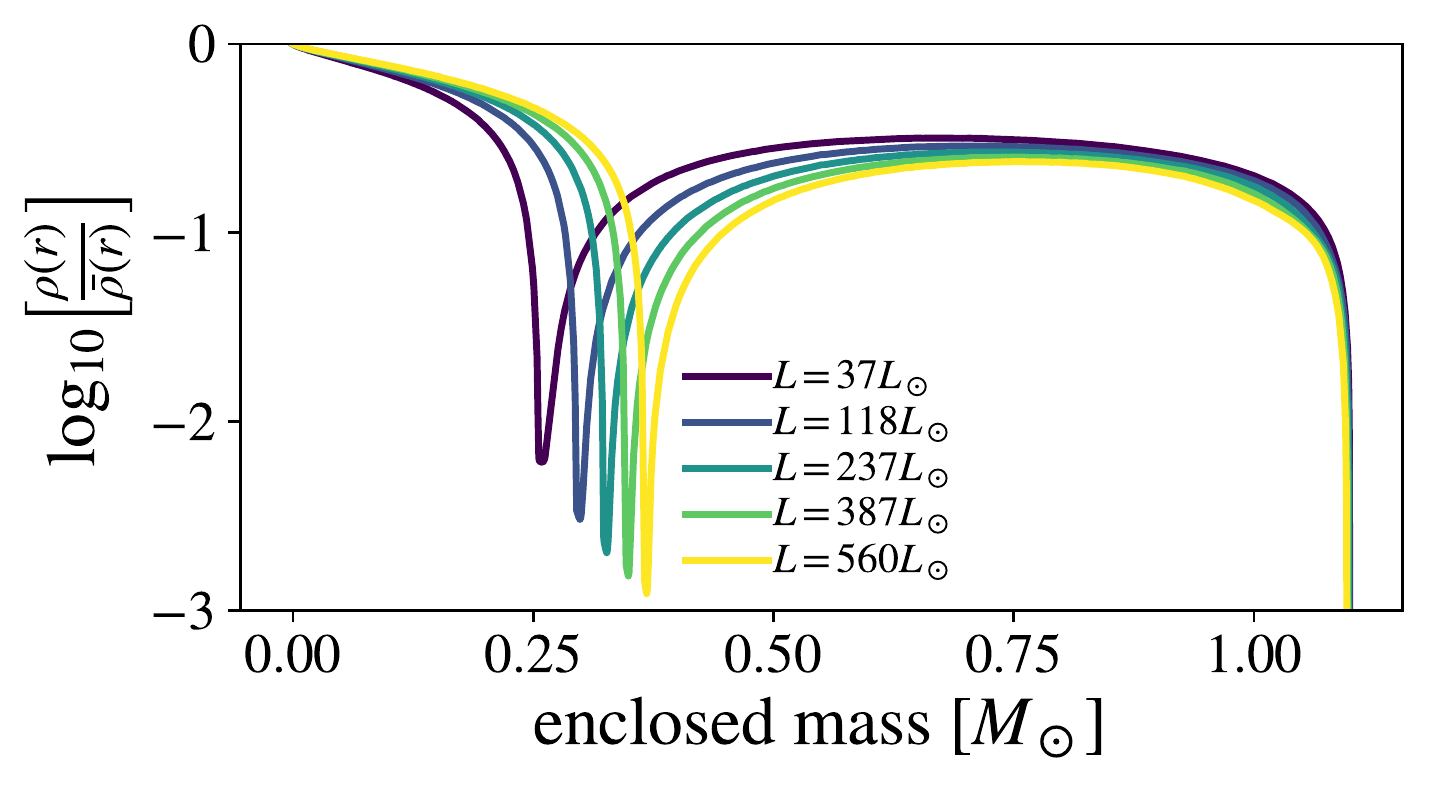}
\caption{Ratio of density $\rho(r)$ to enclosed average density, $\bar \rho(r) = m(r) / (4/3 \pi r^3)$, within the envelope of a 1.1~$M_\odot$ star as it evolves up the giant branch approximately 8~Gyr after zero-age main sequence.  Lines are labeled by stellar luminosity, but growing core mass can also be observed by the dip in $\rho/\bar\rho$. The stellar profiles are computed with MESA version 7503 \citep{2011ApJS..192....3P,2013ApJS..208....4P,2015ApJS..220...15P}, with parameters from Mesa Isochrones and Stellar Tracks \citep{2016ApJ...823..102C}.  We find that $\rho(r)/\bar \rho(r) \approx 0.1$ across much of the giant's envelope, without substantial variation as the star evolves.   }
\label{fig:rho}
\end{center}
\end{figure}

We approximate the density in equation \eqref{drag} as proportional to the average density of the entire star:
\beq
\rho\approx\eta\bar\rho(R_*) = \eta \frac{3 M_*}{4 \pi R_*^3}.
\eeq 
Within the envelope of an evolved star, $\rho < \bar \rho$ (or $\eta<1$) because mass is concentrated in the stellar core.
Figure~\ref{fig:rho} shows a profile $\rho(r) / \bar \rho(r)$ within the interior of a 1.1~$M_\odot$ star as it evolves up the giant branch $(r<R_*)$. Within the envelope a typical $\eta = \rho(r) /\bar \rho(r) \approx 0.1$.

 Combining these ingredients yields 
\beq\label{ldecay}
L_{\rm decay} \approx {3 \over 4} C_{\rm d} \eta G^{3/2} R_2^2 M_* \left( M_* +M_2 \right)^{3/2} R_*^{-9/2}.
\eeq
We adopt $C_{\rm d}=1$ and $\eta=0.1$ in what follows. 
The  number of orbits over which this orbital decay occurs is 
\beq
N_{\rm decay} \approx { E_{\rm orb} \over L_{\rm decay} P_{\rm orb} },
\eeq
which scales as $N_{\rm decay} \propto \left(M_2/ M_*\right) \left( R_*/R_2\right)^2$ for $M_*\gg M_2$.  

The above estimates consider the initial plunge and inspiral of the planetary companion within the stellar envelope. This is reasonable, because in generating an order unity change in the orbital energy, the planet intersects a column of mass similar to its own mass, with the likely result being the erosion and disintegration of the planet  \citep{1998ApJ...506L..65S,2002ApJ...572.1012S}.  If a planet could inspiral to greater depth before being eroded, the larger mean density and higher orbital velocity imply that $L_{\rm decay}$ would exceed our estimate that considers $a\approx R_*$.  

\section{Engulfment In the H--R Diagram}\label{sec:hr}

In this section we examine how the orbital decay power estimated in the previous section compares to the intrinsic luminosity of evolving giant stars. Because $L_{\rm decay}$ decreases as stars evolve up the giant branch while $L_*$ increases, we find that the relative energetics of consuming a planetary companion depend highly on the star's evolutionary state, and therefore its location in the H--R diagram.

To facilitate this comparison, we use stellar data tabulated by the MESA\footnote{http://mesa.sourceforge.net/} Isochrones and Stellar Tracks (MIST) project\footnote{http://waps.cfa.harvard.edu/MIST/} \citep{2016ApJ...823..102C}.  We use isochrones of solar metallicity, non-rotating post-main-sequence stars, which are available for download among MIST's model grids. 

\begin{figure}[tbp]
\begin{center}
\includegraphics[width=0.49\textwidth]{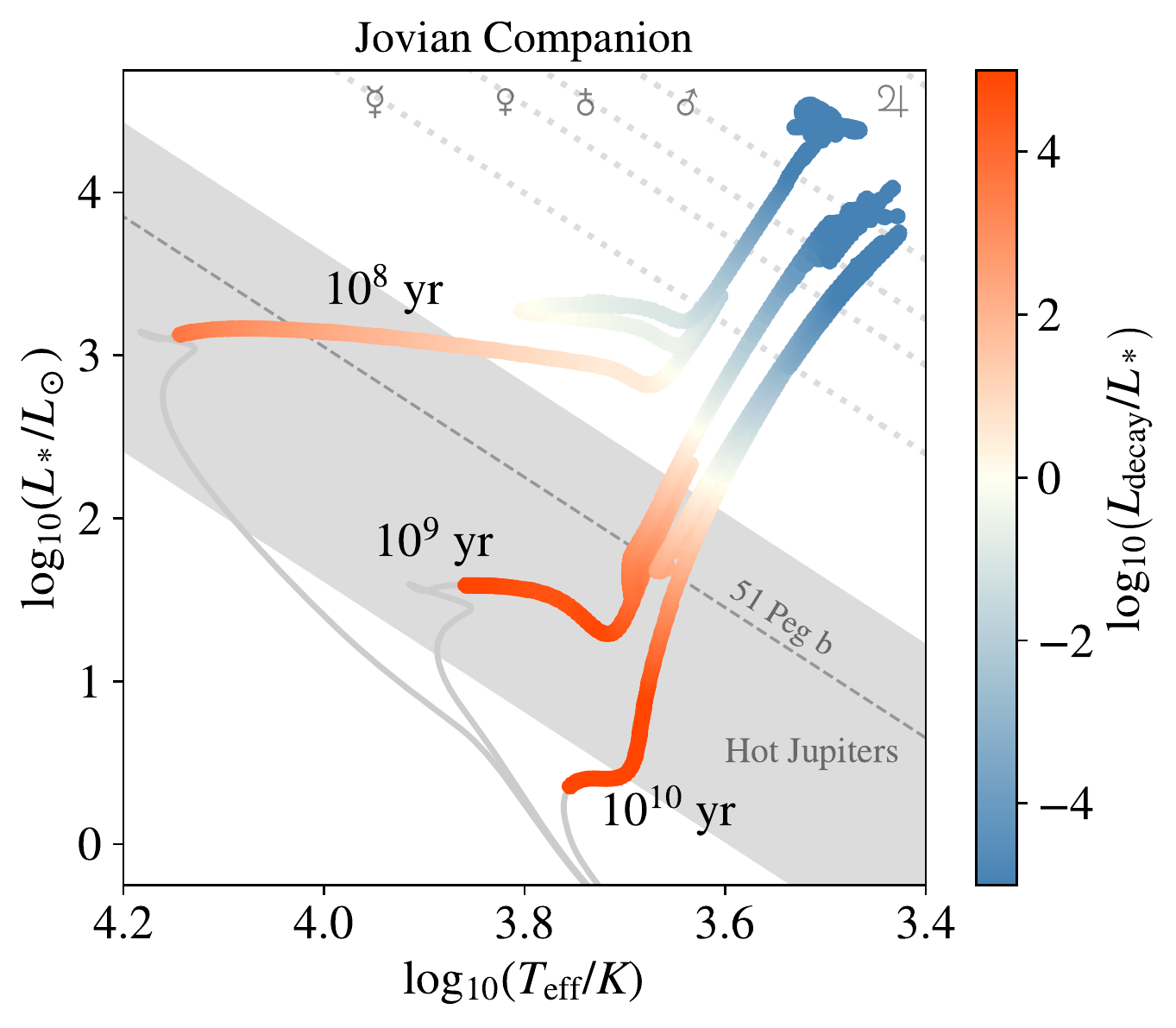}
\caption{Initial orbital decay power, equation \eqref{ldecay}, compared to the intrinsic stellar luminosity. We plot the ratio of $L_{\rm decay}/L$ for solar metallicity, non-rotating stars with isochrone ages $10^8$, $10^9$, and $10^{10}$~yr. In the background the orbital parameters of the inner solar system planets and of hot Jupiters are plotted. A Jovian planet engulfed by a star at the tip of the giant branch (or, equivalently, one at several au separation) has minimal impact on the power output of its host star. In the opposite case, a close-in Jovian planet, such as a hot Jupiter, will deposit orders of magnitude more power in orbital decay than the intrinsic luminosity of a compact host star near the giant-branch turnoff.   }
\label{fig:LoL}
\end{center}
\end{figure}

We begin by examining the ratio of orbital decay power generated by the engulfment of a Jovian giant planet ($R_2 =6.99\times10^9$~cm) to intrinsic stellar luminosity.  
Figure \ref{fig:LoL} shows this ratio along post-main-sequence isochrones of $10^8$, $10^9$, and $10^{10}$~yr age in the H--R diagram.  The ratio of orbital decay power to stellar luminosity spans nearly ten orders of magnitude across the H--R Diagram. We find that near the main-sequence turnoff, where orbits are compact, the associated orbital decay timescale is short, and the power injected through digestion of a giant planet is large compared to the nuclear-burning luminosity of the star. At the tip of the giant branch, the opposite is true and the initial orbital decay of a giant planet within the envelope would result in a small perturbation to the intrinsic stellar luminosity.

Conditions where orbital decay power and stellar luminosity are similar mark a critical transition between stars that could be largely undisturbed by swallowing a giant-planet companion to those  whose structure would be dramatically modified. For stars of $10^{10}$~yr age, the transition occurs about halfway up the giant branch around luminosities of 100~$L_\odot$. 

Locations in the $L_*$, $T_{\rm eff}$ axes of the H--R diagram map to a radius through an assumption of blackbody emission, $R_*^2=L_*/4\pi\sigma T_{\rm eff}^4$. These radii can be compared to the orbital separations of populations of planets.  Figure \ref{fig:LoL} reveals that when a star engulfs a hot Jupiter companion (with, for example, $a\approx0.05$~au), we expect a dramatic imprint on the host star, with $L_{\rm decay}$ hundreds of times $L_*$. On the other hand, the engulfment of a typical warm Jupiter (with $0.5$~au separation) will result in a minor disturbance to a typical host star with $L_{\rm decay}$ only at the percent level of $L_*$.

\begin{figure}[tbp]
\begin{center}
\includegraphics[width=0.49\textwidth]{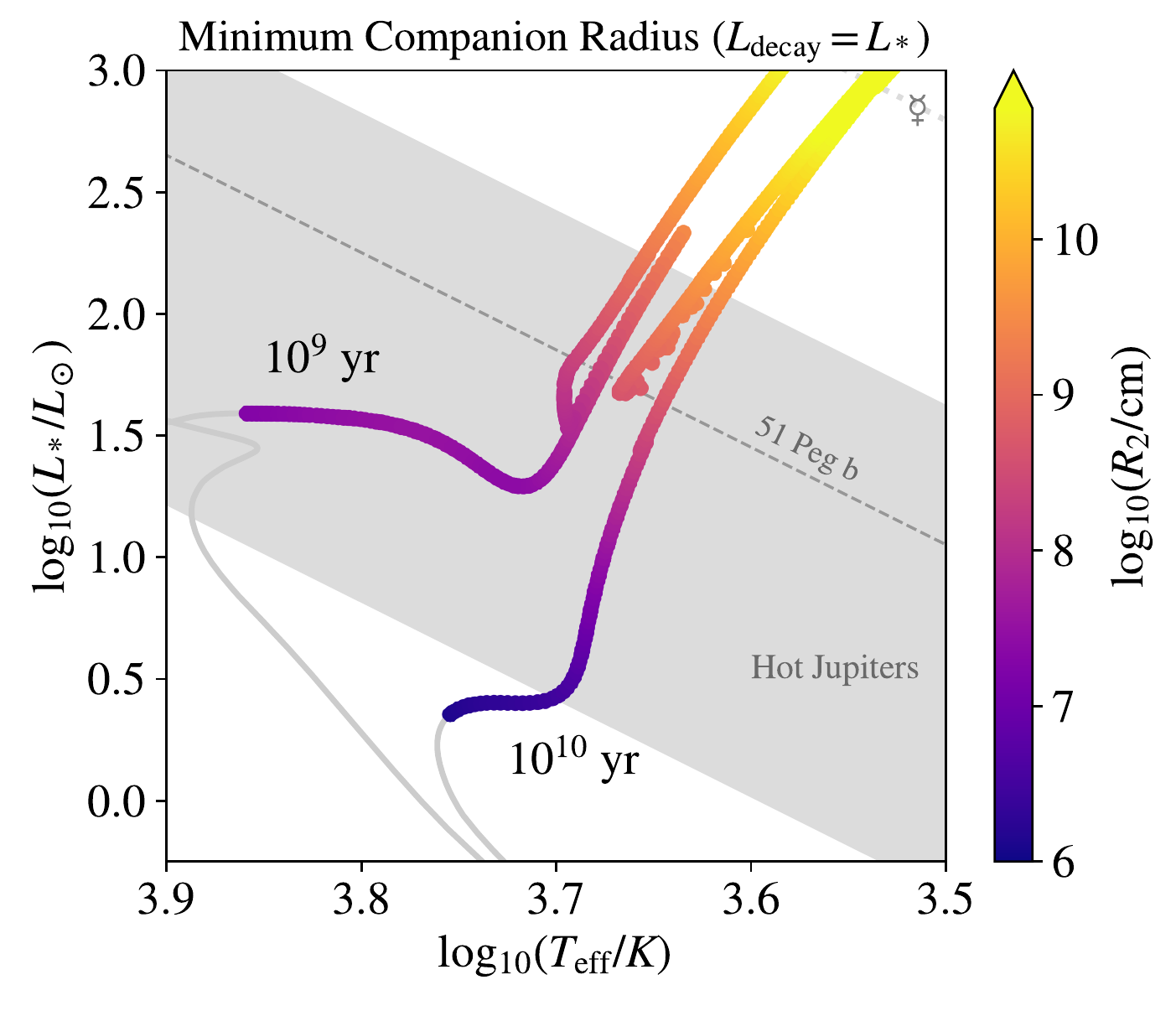}
\caption{Minimum radius of a companion whose initial orbital decay power matches the intrinsic stellar luminosity,  $L_{\rm decay}=L_*$.  Here, we focus on the lower giant branch of $10^9$ and $10^{10}$~yr isochrones. Near the main-sequence turnoff, objects as small as 10~km radius can have $L_{\rm decay}\approx L_*$. The critical radius quickly grows to roughly an Earth radius at hundredths of an au and Jupiter radius at 0.1~au. } 
\label{fig:R2}
\end{center}
\end{figure}

In Figure \ref{fig:R2}, we extend this consideration of the critical condition. We compute the minimum radius companion  that yields $L_{\rm decay} = L$. The minimum radius is only about $10$~km near the main sequence turnoff, but quickly grows to the size of low-mass rocky planets at hundredths of an au. By $a\approx 0.1$~au, the critical radius is similar to the radius of Jupiter, as can be inferred from Figure \ref{fig:LoL}. Going to the tip of the giant branch the critical radius becomes so large (greater than $R_\odot$) that the object in question would be stellar-mass rather than planetary-mass.  Under those circumstances, gravitational focusing dominates the cross section as discussed in Section \ref{sec:orbdecay}, and it is the mass, rather than the radius, of the companion which is constrained.  

\begin{figure*}[tbp]
\begin{center}
\includegraphics[width=0.49\textwidth]{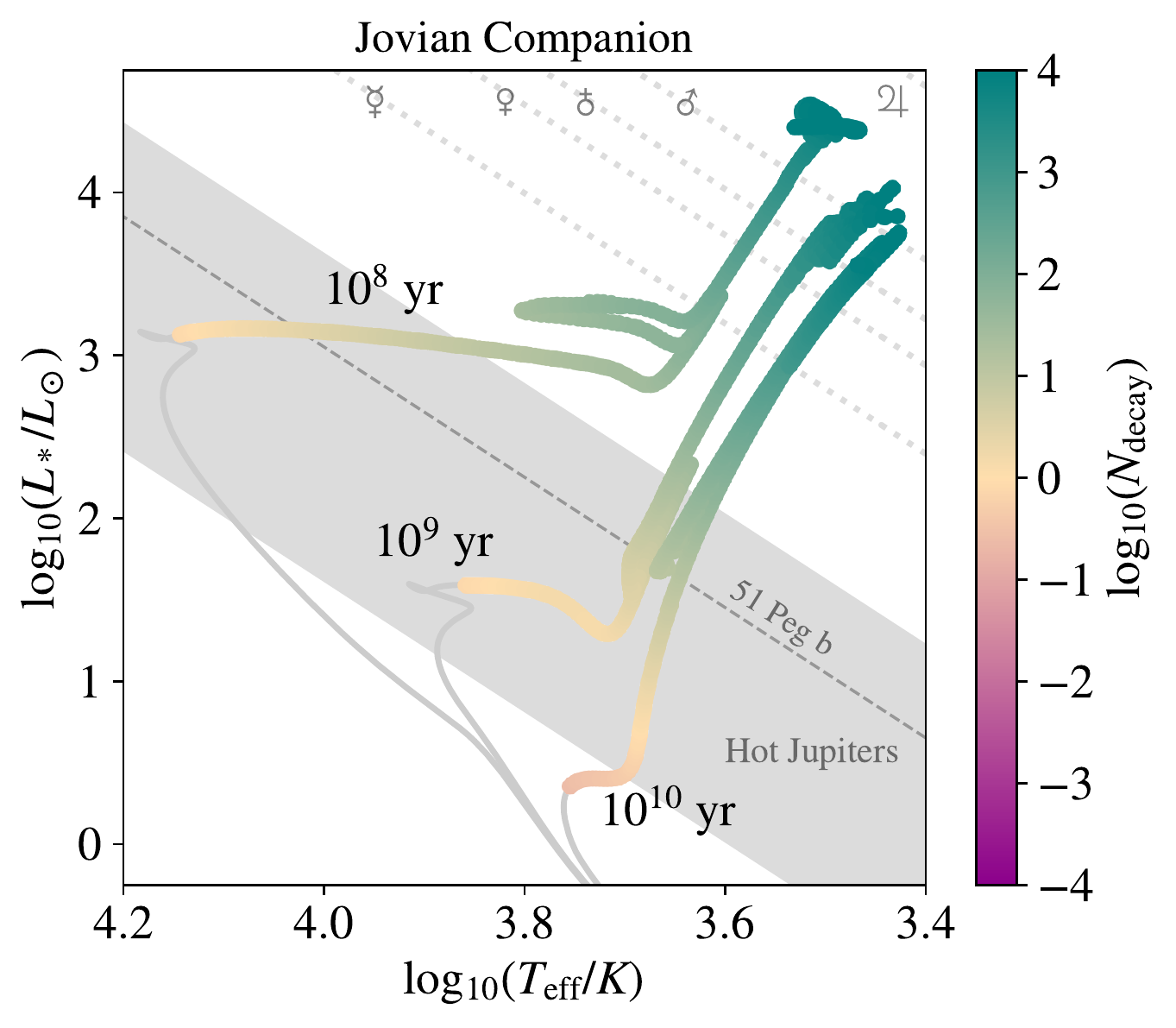}
\includegraphics[width=0.49\textwidth]{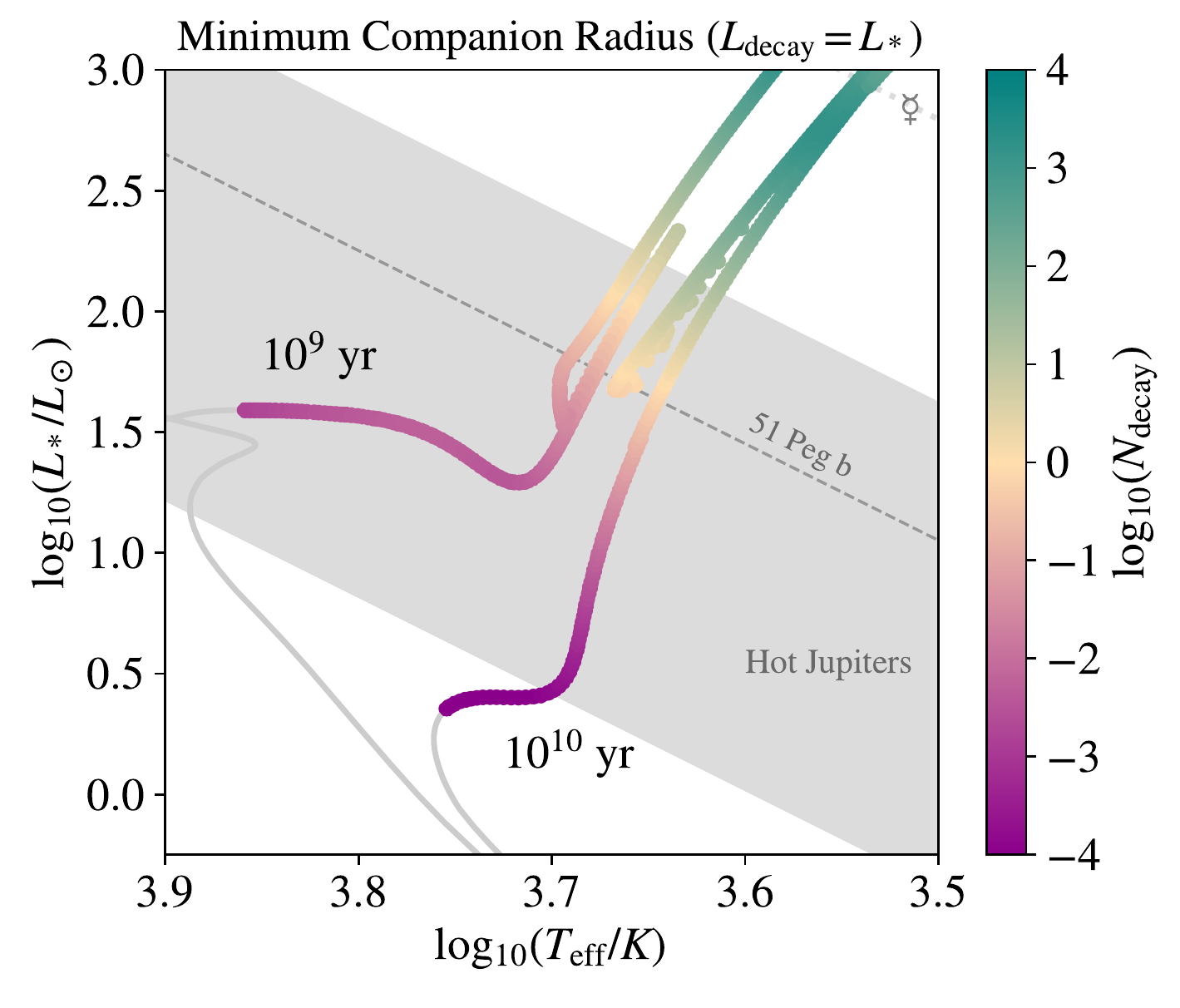}
\caption{ Number of orbits over which engulfed planet's orbits decay. The left panel considers a Jovian planet, while the right panel considers an object with the minimum radius of Figure \ref{fig:R2} and Jovian density.  Higher up the giant branch, it takes many orbits for an object's orbit to decay. Near the turnoff, it takes of order one orbit for an engulfed Jovian planet's orbit to decay and a minimum size engulfed body's orbit decays in a tiny fraction of an orbital period. This indicates that these small engulfed objects survive only briefly within their host star's atmosphere.} 
\label{fig:tdecay}
\end{center}
\end{figure*}

Finally, we examine the number of orbital periods over which an engulfed planet's orbit decays. The left panel of Figure \ref{fig:tdecay} plots $N_{\rm decay}$ for engulfed Jovian planets; the right panel considers the minimum radius object ($L_{\rm decay} =L_*$) of Figure \ref{fig:R2}. Jovian planets undergo orbital decay (and, as noted earlier, likely ablation on a similar timescale) over about $1$ orbit near the giant-branch turnoff to about $10^4$ orbits near the giant-branch tip. This highlights, in part, why $L_{\rm decay}/L_*$ is so varied across the H--R diagram, the number of orbits over which the orbital energy is dissipated varies significantly. Minimum radius objects can have extraordinarily small $N_{\rm decay}\approx 10^{-4}$ near the giant-branch turnoff. This indicates that the net injected energy (approximately $E_{\rm orb}$) is small, and that, in some cases, these objects are so vulnerable that they likely would not survive interaction with a host star's extended atmosphere to become truly engulfed. 

\section{Discussion}\label{sec:discussion}

 Having established the comparison of $L_{\rm decay}$ to $L_*$ we discuss its interpretation below.

\subsection{Emergent Luminosity versus Orbital Decay Power}

Because orbital decay power is generally deposited into very optically thick stellar envelope material,  the emergent luminosity is mediated by transport to the stellar photosphere. 
Nonetheless, orbital decay power can give rise to both short- and long-term electromagnetic emission. 

Upon engulfment of a companion, immediate signatures may include cooling emission from shock-heated ejecta \citep{2012MNRAS.425.2778M,2017MNRAS.466.1421Y}, and  radio afterglows from ejecta--interstellar medium interactions \citep{2017MNRAS.466.1421Y}.
On a dynamical timescale the stellar envelope expands to return to quasi-hydrostatic equilibrium as orbital decay  does work against the gravitational binding energy of the envelope. 
 \citet{2016MNRAS.458..832S} model the interaction of RGB and AGB stars near the middle and tip of the giant branch with a 10 Jupiter-mass planet and find mild hydrostatic expansion of the objects post-engulfment (along with some, marginally resolved, ejecta). 
Because these simulations do not model photospheric cooling, this expansion reflects the return to a perturbed hydrostatic equilibrium following the deposition of orbital decay energy.

Longer-term signatures include the eventual transport of heat deposited within layers of the envelope to the stellar surface.  The emergent luminosity in this phase is $\Delta L \approx\Delta E(r) / t_{\rm KH}(r)$, where $\Delta E(r)$ is the locally deposited energy and $ t_{\rm KH}(r)$ is the local Kelvin--Helmholtz, or transport, timescale (by convection or radiative diffusion) for that energy  through the perturbed stellar envelope structure. 

 When the orbital decay timescale is short compared to that of energy transport, the initial response of the stellar envelope is expansion, and luminosity emerges only on the longer timescale of energy transport. This implies that cases where $L_{\rm decay} \gg L_*$ will have net emergent luminosity lower than $L_{\rm decay}$ but higher than $L_*$. When the timescale for energy transport is short compared to deposition, radiative cooling from the photosphere moderates the extra energy from orbital decay. For example, if cooling were included in the AGB model of  \citet{2016MNRAS.458..832S}, radiation from the photosphere would carry away energy at a similar rate to deposition.

Finally, we note that under the simplifying  assumption that orbital decay energy is distributed spherically, the quasi-hydrostatic aspect of energy transport can be addressed in 1D simulations with stellar evolution codes \citep[see, for example, recent calculations presented by][]{2017MNRAS.468.4399M}.

\subsection{Engulfment and Stellar Rotation}

In addition to energy, orbital decay also deposits angular momentum into stellar envelope material.  Prior to engulfment, a planet's orbit may decay through tidal interaction with the host star \citep[e.g.][]{1977A&A....57..383Z,1980A&A....92..167H,1996ApJ...470.1187R}, in which case much of the angular momentum is expected to be deposited into the host star's convective zone \citep{2014ApJ...786..102V,2017A&A...602L...7M}. 
Stellar evolution calculations including tidal interactions and the uncertain physics of internal angular momentum transport \citep{2016A&A...593A.128P}
support the notion that planet engulfment could be responsible for some of the observed fast rotating red giants \citep{2009ApJ...700..832C,2012ApJ...757..109C}. 
The distribution of angular momentum deposition following engulfment is considerably less certain \citep[though see the characterization of][for their simulated cases]{2016MNRAS.458..832S}. 

A related signature of planetary engulfment that is often considered together with enhanced rotation is  anomalous surface abundances, in particular Lithium enrichment \citep[e.g.,][]{1967Obs....87..238A,1999MNRAS.308.1133S}.
However,  due to the fragile nature of Lithium,  this  particular signature is likely short-lived and degenerate with other mixing processes \citep{2016A&A...593A.128P}.  

\subsection{Companion Engulfment Rates}
To estimate the rate at which planets are engulfed as their host stars ascend the giant branch, we consider the occurrence rate of close-in planets and the rate at which stars evolve off of the main sequence. The current rate at which $1M_\odot$ stars become giants reflects the Milky Way star formation history of approximately $10^{10}$~yr, a main-sequence lifetime, ago. 
 The star formation rate of Milky Way progenitor stars $10^{10}$~yr ago, at approximately redshift 2, has been inferred to be approximately 10~$M_\odot$~yr$^{-1}$ \citep{2013ApJ...771L..35V}. 
Planet occurrence rates vary by planetary and stellar type. Among the Kepler sample, approximately 10\% of stars host giant planets, while roughly $1$\% host close-in hot Jupiters \citep[e.g.][]{2015ARA&A..53..409W}.
 The rate of engulfment in the Milky Way can then be written
\beq
R_{\rm engulf} \approx f_{\rm comp} R_{\rm evol}
\eeq
where $ f_{\rm comp} $ is the companion occurrence fraction, and $R_{\rm evol}$ is the rate at which stars evolve off the main sequence. For the typical parameters listed above, we therefore find $R_{\rm engulf} \approx 0.1$--$1$~yr$^{-1}$ in the Galaxy depending on what type of companion is considered \citep[see][for a more detailed discussion including the possibility of planet migration]{2012MNRAS.425.2778M}. We note that this rate is similar to the inferred Galactic rate of  stellar mergers  \citep{Kochanek2014}.

\section{Summary}\label{sec:conclusion}
Planets may be engulfed by their evolving host stars when these stars expand on the giant branch. We have compared the power from planetary orbital decay to stellar luminosity across the H--R diagram. Some key considerations are as follows. 

\begin{enumerate}

\item Drag forces cause the orbit of an engulfed planet to decay, as estimated in equations \eqref{drag} and \eqref{ldecay}, respectively.

\item Whether or not the engulfment of a planet will significantly perturb a giant-branch star's energetics depends on the properties of the planet (mass, separation) and on the stellar luminosity when it evolves to $R_*\approx a$.  The orbital decay power falls steeply with increasing stellar size because the stellar envelope density goes down (roughly $L_{\rm decay} \propto R_*^{-9/2}$), while luminosity grows sharply with radius in low-mass giants, roughly  $L_* \propto R^{3/2}$ \citep{1987ApJ...319..180J}.  These opposed scalings indicate the presence of a crossing, where in some conditions $L_{\rm decay}> L_*$ and in others $L_{\rm decay}< L_*$. 

\item Giant planets on orbits with $a\lesssim 0.1$~au give rise to  $L_{\rm decay} \gg L_*$. Those on wider orbits have $L_{\rm decay}\ll L_*$ (Figure \ref{fig:LoL}). This indicates that hot Jupiters substantially affect their host stars upon engulfment, while warm Jupiters ($a\approx 0.1$--$1$~au) do not. Evolved stars of $10^{9}$ to $10^{10}$~yr age and $L\approx 10^2$ ~$L_{\odot}$ represent the transition population where $L_{\rm decay}\approx L_*$ for a Jovian planet. 

\item We compute the radius of objects that match the critical condition $L_{\rm decay}=L_*$. These are as small as $R_2 \approx 10$~km for host stars at the giant-branch turnoff. 

\item The timescale of energy injection varies from of order a single orbital period for a Jovian planet engulfed by a host near the turnoff, to $10^4$ orbital periods at the tip of the giant branch. Minimum radius objects are short-lived with $N_{\rm decay}\ll 1$. 

 \item Emergent luminosity will be mediated by transport through the optically thick stellar envelope and is not expected to match $L_{\rm decay}$, particularly when $L_{\rm decay}>L_*$.  

\end{enumerate}

While the engulfment of planets by their host stars is a common event in that many to most stars are subject to such an interaction during their lifetime, the present-day specific event rate is low, of the order of 0.1--1~yr$^{-1}$ in the Galaxy. This fact motivates examination of the long-lasting observational signatures of companion interaction with a focus on stars ascending the lower giant branch with $L_*< 100$~$ L_\odot$.

\section*{Acknowledgements }
We gratefully acknowledge thoughtful and detailed feedback from the referee, C. Tout, and helpful discussions with  A. Antoni, J. Grindlay, A. Loeb, C. Morley, K. Masuda, E. Ostriker, E. Ramirez-Ruiz, J. Stone, and S. Tremaine. Support for this work was provided by NASA through Einstein Postdoctoral Fellowship grant number PF6-170169  awarded by the Chandra X-ray Center, which is operated by the Smithsonian Astrophysical Observatory for NASA under contract NAS8-03060.

\bibliographystyle{aasjournal}

\end{document}